\documentclass[sigconf]{acmart}
\AtBeginDocument{%
  }

\setcopyright{acmlicensed}
\usepackage{listings}
\usepackage{hyperref}
\usepackage{xurl} 
\usepackage{siunitx}
\usepackage{float}
\usepackage{amsmath} 
\usepackage{array} 
\usepackage{geometry} 
\usepackage{longtable} 
\usepackage{xcolor} 
\usepackage{arydshln}
\usepackage{xspace}
\newcommand{\tool}{\textsc{ETrace}\xspace}

\usepackage{caption}
\usepackage{longtable} 

\title{\tool:Event-Driven Vulnerability Detection in Smart Contracts via LLM-Based Trace Analysis}

\author{%
  Chenyang Peng$^{1}$,
  Haijun Wang$^{1}$,
  Yin Wu$^{1}$,
  Hao Wu$^{1}$,
  Ming Fan$^{1}$,
  Yitao Zhao$^{2}$,
  Ting Liu$^{1}$
}

\affiliation{%
  \institution{Xi’an Jiaotong University, Xi’an, Shaanxi}
  \country{China} 
}

\affiliation{%
  \institution{Yunnan Power Grid Co., Ltd, Yunnan}
  \country{China}
}
\acmConference[Internetware: New Idea'25]{16th Asia-Pacific Symposium on Internetware: New Idea}{June 20-June 22, 2025}{Trondheim, Norway}
\acmBooktitle{16th Asia-Pacific Symposium on Internetware: New Idea, June 20-June 22, 2025, Trondheim, Norway}

\acmISBN{xxx-x-xxxx-xxxx-x/xxxx/xx}
\acmDOI{10.1145/3755881.3755934}

\begin{abstract}

With the advance application of blockchain technology in various fields, ensuring the security and stability of smart contracts has emerged as a critical challenge.
Current security analysis methodologies in vulnerability detection can be categorized into static analysis and dynamic analysis methods.
However, these existing traditional vulnerability detection methods predominantly rely on analyzing original contract code, not all smart contracts provide accessible code.
We present ETrace, a novel event-driven vulnerability detection framework for smart contracts, which uniquely identifies potential vulnerabilities through LLM-powered trace analysis without requiring source code access. By extracting fine-grained event sequences from transaction logs, the framework leverages Large Language Models (LLMs) as adaptive semantic interpreters to reconstruct event analysis through chain-of-thought reasoning. ETrace implements pattern-matching to establish causal links between transaction behavior patterns and known attack behaviors.
Furthermore, we validate the effectiveness of ETrace through preliminary experimental results.
\end{abstract}



\received{5 April 2025}            
\received[revised]{10 May 2025}   
\received[accepted]{20 April 2025}  

\begin{document}

\maketitle

\section{Introduction}

In recent years, blockchain technology has garnered widespread attention across various domains due to its characteristics of immutability, trustworthiness, and decentralization.
While decentralized exchanges (DEXs)\cite{2} exhibit functional diversity across automated market making (e.g., Uniswap) and hybrid architectures, their rapid proliferation has made them prime targets for malicious actors\cite{3} and scam tokens\cite{wu2024tokenscout}.
In the NewFreeDAO incident($\$$125M loss), attacker exploited a flash loan to purchase a large amount of NFD tokens, substantial rewards, and then repaid the flash loan. In the Omni NFT ($\$$1.4M loss) hack exploits the reentrant vulnerability to hijack the contract execution process.

Current research primarily relies on code analysis tools (e.g., Slither, MythX) to detect vulnerabilities. However, most real-world DEX attacks involve third-party dependencies (e.g., oracle feeds, bridge contracts) whose source code is inaccessible. 

To process this obstacle, we shift our focus to transaction logs.
Due to the decentralized nature of blockchain and its tamper-resistant properties, it has garnered significant popularity. As a result, the trading volume of DEXs continues to grow steadily. As DEX trading volumes grow daily, each transaction is accompanied by the generation of transaction logs.

These logs capture event signatures that reveal both explicit invocation dependencies and implicit interaction patterns among smart contracts.
Previous research has extensively explored the event information in transaction logs. For example, Kaleem et al.\cite{4} utilized event information as the core communication mechanism to trigger contract logic, reducing the average total delay of event-triggered smart contracts by 2.2 to 4.6 times. Liu et al.\cite{5} conducted an in-depth analysis of phantom event-related issues, addressing problems such as event forgery, inconsistent records, and contract impersonation. Meanwhile, Salzano et al.\cite{6} employed logs to test smart contracts, aiming to compare log outputs with expected results. This demonstrates that, in the current context of frequent attack incidents, the event information of smart contracts can also provide a new perspective for vulnerability detection.
However, such deficiency between event interpretation and attack causality severely limits detection accuracy.

To address these challenges, we propose ETrace, a framework that leverages event information extracted from transaction logs to infer the intent of smart contract transactions using LLMs.

\begin{figure*}[h]
\vspace{-2em}
    \centering
    \includegraphics[width=1\textwidth]{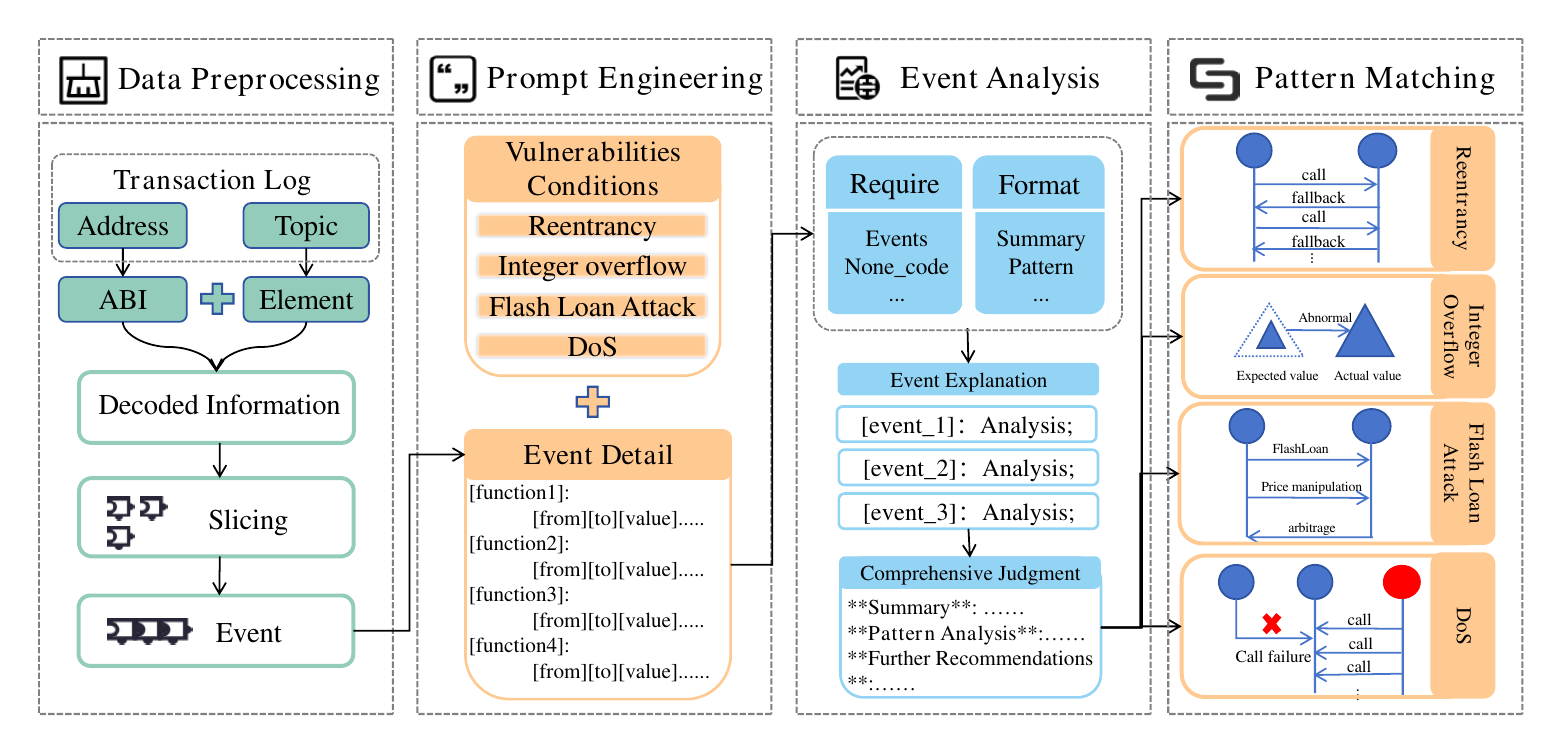}
    \vspace{-3em}
    \caption{\centering Architecture for ETrace.}
    \vspace{-2em}
    \label{fig:pdf}
\end{figure*}

As illustrated in Figure~\ref{fig:pdf}, ETrace first retrieves event information from smart contract transaction logs.
Then, LLMs are employed as experts, processing characteristics of attack behaviors to analyze each event individually which mines the relationship between transaction behaviors and attack behaviors.
Finally, events are compared with four predefined attack patterns through pattern matching to identify potential vulnerabilities in smart contracts.
Our preliminary experiments have confirmed the effectiveness of ETrace. The main contributions of this paper are as follows:\vspace{-1.6em}
\begin{itemize}
\item By utilizing large models to process features from events extracted by transaction logs and perform semantic intent analysis, we achieve a more precise understanding of the transaction behaviors of smart contracts.

\item We propose a vulnerability detection framework that utilizes LLMs to semantically analyze events without source code, identifying four attack types: reentrancy, integer overflow, flash loan attacks, and DoS.

\item Experiments are conducted to utilize a dataset of four real-world attack incidents, and the results successfully validate the effectiveness of ETrace.

\end{itemize}

\section{Motivation example}

Through the study of event information from different attack incidents, we observed that the event characteristics of the following types of attacks exhibit high distinctiveness.

\begin{table}[ht]
\vspace{-1em}
\caption{Reentrancy Event}
\vspace{-1em}
\centering
    \begin{tabular}{@{}p{2cm}lp{2cm}lp{3cm}l@{}}
        \toprule
        \textbf{Event} & \textbf{Address} & \textbf{Value} \\
        \midrule
        Transfer & 0x0eD7$\to$0x5f2e & \num{1.0000e+22} \\
        Transfer & \textbf{0xE1E1}$\to$\textbf{0x5f2e} & \num{1.8966e+15} \\
        Transfer & \textbf{0xE1E1}$\to$\textbf{0x5f2e} & \num{1.1235e+16} \\
        Transfer & \textbf{0x5f2e}$\to$\textbf{0xE1E1}& \num{1.8966e+15} \\
        Transfer & \textbf{0xE1E1}$\to$\textbf{0x5f2e}& \num{1.2964e+18} \\
        Transfer & 0x5f2e$\to$0xE1E1 & \num{1.1235e+16} \\
        Transfer & 0xE1E1$\to$0x5f2e & \num{1.8641e+18} \\
        Transfer & 0x5f2e$\to$0xE1E1 & \num{1.2964e+18} \\
        Transfer & 0x5f2e$\to$0xE1E1 & \num{1.8641e+18} \\
        Transfer & 0x5f2e$\to$0x0eD7 & \num{1.0030e+22} \\
        \bottomrule
    \end{tabular}
\label{reentrancy}
\end{table}

As shown in Table \ref{reentrancy}, a reentrancy attack occurred on XSURGE\footnote{\url{https://medium.com/@Knownsec_Blockchain_Lab/knownsec-blockchain-lab-comprehensive-analysis-of-xsurge-attacks-c83d238fbc55}} (Aug 15, 2021), where attacker 0x5f2e repeatedly invoked the victim contract (0xE1E1)’s sell and purchase functions.  These emitted consecutive Transfer events, with alternating from/to addresses between 0x5f2e and 0xE1E1, forming a loop that enabled the exploit.



\begin{table}[ht]
\vspace{-2em}
\caption{Integer Overflow}
\vspace{-1em}
\centering
    \begin{tabular}{@{}p{2cm}lp{2cm}lp{3cm}l@{}}
        \toprule
        \textbf{Event} & \textbf{Address} & \textbf{Value} \\
        \midrule
        Transfer & 0x09a3$\to$0xb4D3 & $2^{255}$ \\
        Transfer & 0x09a3$\to$0x0e82 & $2^{255}$ \\
        \bottomrule
    \end{tabular}

\label{overflow}
\end{table}

\begin{sloppypar}
Table \ref{overflow} is an integer overflow incident that occurred on Beauty Chain\footnote{\url{https://etherscan.io/tx/0xad89ff16fd1ebe3a0a7cf4ed282302c06626c1af33221ebe0d3a470aba4a660f}}
 on April 22, 2018.  Due to the input parameter exceeded the range of an unsigned integer, an integer overflow occurred, causing the value field to reach an unusually large magnitude (specifically, $2^{255}$).  This allowed the attacker to profit by 900 million dollars.  The event characteristic of this case is clearly reflected in the abnormal value field.
\end{sloppypar}

\begin{table}[ht]
\caption{Flash Loan Attack Event}
\vspace{-10pt}
\centering
\begin{tabular}{@{}p{2cm}lp{2cm}lp{5cm}l@{}}

\toprule
\textbf{Event} & \textbf{Address} & \textbf{Value} \\
\midrule
Transfer & 0x2D00$\to$0x88e6 & $1.8774\, \times \! 10^{20}$ \\
Transfer & 0x0000$\to$0xBA12 & $1.0000\, \times \! 10^{0}$ \\
\textbf{FlashLoan} & \textbf{0x0000$\to$0xC02a} & \textbf{$1.0000\, \times \! 10^{0}$} \\
Transfer & 0x8ad5$\to$0x4b77 & $1.5825\, \times \! 10^{20}$ \\
Transfer & 0xB4e1$\to$0x4b77 & $1.6941\, \times \! 10^{19}$ \\
Approval & 0x4b77$\to$0xBA12 & $1.1579\, \times \! 10^{77}$ \\
Swap & None$\to$None & out0=0, out1=0 \\
Transfer & 0x4b77$\to$0xBA12 & $1.4707\, \times \! 10^{10}$ \\
Transfer & 0xBA12$\to$0x4b77 & $1.1379\, \times \! 10^{19}$ \\
Transfer & 0x4b77$\to$0xB4e1 & $2.1896\, \times \! 10^{10}$ \\
Transfer & 0x4b77$\to$0x8ad5 & $2.0454\, \times \! 10^{11}$ \\
Sync & & in=$4.6287\, \times \! 10^{13}$ \newline \text{$out=3.5902\, \times \! 10^{22}$} \\
\textbf{Swap} & \textbf{0x4b77} & \text{in=$2.1896\,\times \! 10^{10}$},\newline \text{out=$1.6941\, \times \! 10^{19}$} \\
\textbf{Swap} & \textbf{0x4b77} & \text{in=$2.0454\,\times \! 10^{11}$},\newline 
out=0 \\
\textbf{Withdrawal} & \textbf{0x4b77} & \textbf{$1.8657\,\times \! 10^{20}$} \\
\bottomrule
\end{tabular}

\label{flash}
\end{table}
\begin{sloppypar}

Table~\ref{flash} presents a flash loan attack on MEVBOT (Oct 14, 2022)\footnote{\url{https://etherscan.io/tx/0x35ecf595864400696853c53edf3e3d60096639b6071cadea6076c9c6ceb921c1}}. Although MEVBOT lacks source code, event data shows the attacker transferred \( 1.8774 \times 10^{20} \) to~0x88e6, conducted cyclic trades via~0xB4e1 and~0x8ad5, and authorized \( 1.1579 \times 10^{77} \) tokens to~0xBA12 for arbitrage. Through multiple \texttt{Swap} operations, they gained profit and withdrew \( 1.8657 \times 10^{20} \) via a \texttt{Withdrawal} event.

\end{sloppypar}


\begin{table}[ht]
\caption{DoS Event}
\vspace{-10pt}
\centering
\begin{tabular}{@{} l l c r @{}}  
\toprule
\textbf{Event} & \textbf{Address} & \textbf{Value} & \textbf{Gas} \\  
\midrule
\textbf{lendGM} & 0x94bd$\to$0xf457 & 0.01 & \textbf{36855} \\  
totalPayedOut() & 0x490f$\to$0xf457 & 0.0 & 21651 \\  
\textbf{lendGM} & 0x818d$\to$0xf457 & 0.001 & \textbf{2532963} \\  
\textbf{lendGM} & 0x818d$\to$0xf457 & 0.001 & \textbf{5057945} \\  
\textbf{lendGM} & 0x818d$\to$0xf457 & 0.001 & \textbf{5057945} \\  
\textbf{lendGM} & 0x818d$\to$0xf457 & 0.001 & \textbf{5057945} \\  
\textbf{lendGM} & 0x818d$\to$0xf457 & 0.001 & \textbf{5057945} \\  
Unknown Function & 0x490f$\to$0xf457 & 1.0 & 750000 \\  
Unknown Function & 0x490f$\to$0xf457 & 0.8236 & 750000 \\  
Unknown Function & 0x490f$\to$0xf457 & 0.01 & 750000 \\  
\bottomrule
\end{tabular}

\label{dos}
\end{table}

\begin{sloppypar}

Table~\ref{dos} shows a DoS attack on the GovernMental\footnote{\url{https://www.reddit.com/r/ethereum/comments/4ghzhv/governmentals_1100_eth_jackpot_payout_is_stuck/}} contract. To claim a prize, the contract had to delete a creditor list, but due to repeated small ETH loans by malicious user 0x818d via \texttt{lendGM}, the list grew excessively. Deletion required 5{,}057{,}945 gas, exceeding the transaction limit of 4{,}712{,}388, causing 1{,}100 ETH to be permanently locked. Event logs confirm the attacker’s repeated borrowing inflated gas cost and triggered the failure.

\end{sloppypar}

\section{METHODOLOGY}
This module provides a comprehensive overview of the processes and key technologies involved in ETrace.   As illustrated in Figure \ref{fig:pdf}, the architecture of ETrace is divided into four main stages. The process begins with Data Preprocessing, where transaction logs are extracted and transformed into Decoded Information, which is then sliced into events to ensure its suitability for analysis. Next, during the Prompt Engineering phase, the events are enriched with vulnerability conditions and passed to the Event Analysis stage. Here, the events are systematically analyzed, and a Comprehensive Judgment is made based on the event content to align with the required format and conditions. Finally, in the Pattern Matching phase, the analyzed data is evaluated against known vulnerability patterns to identify potential security risks associated with the current event.

\subsection{Data Preprocessing}

We retrieve the transaction log through the transaction hash, which typically contains the contract addresses and topic information for all emitted events. The topics include critical details such as the function signature, sender address, and receiver address within the transaction. By decoding the parameters in the topics using the ABI of the corresponding contract address, we obtain all function names and their associated parameters. The decoded information is then segmented based on the function names, ultimately yielding fine-grained event information.
We categorize event information, into three main components: function name, address, and value. The function name indicates which functions were called between contracts, while the address reveals how many contracts participated in the current behavior.   The value reflects the quantitative measure of resources involved, such as tokens or ETH, associated with the behavior. 

\subsection{Prompt Engineering}
In the prompt engineering phase, we integrate vulnerability conditions with event details.  The vulnerability conditions encompass the characteristics of four known attack types: reentrancy, integer overflow, flash loan attack, and DoS. Taking reentrancy as an example, the vulnerability condition includes ``\textit{multiple calls to the same function during the contract execution, exploiting external calls before state updates, leading to malicious repeated fund transfers or tampering with the contract's state.}'' Additionally, we employ the COT strategy in the prompts, instructing the LLMs to first explain the behavior associated with each event and explicitly output the reasoning process.  Subsequently, the model evaluates the provided explanations holistically to generate a comprehensive judgment.


\subsection{Event Analysis}
In the event analysis phase, we specify the requirements and output format for LLMs to analyze events.   Since the event analysis is conducted without relying on source code, we explicitly include a ``None code'' requirement .  The initial output provides an explanation of each event, which aids in understanding the behavior of smart contracts.   By interpreting these behaviors, we gain deeper insights into the details of how vulnerabilities occur.   The subsequent output builds on the first, delivering a comprehensive evaluation of all event explanation.   Ultimately, the model outputs a structured result in the format of ``Summary - Pattern Analysis - Further Recommendation'', identifying which specific vulnerabilities (or combinations thereof) the event matches.   This step process ensures both granular and holistic analysis, enhancing the accuracy and interpretability of the results.

\subsection{Pattern Matching}
The pattern matching phase serves as a judgment to the pattern analysis results.  To achieve more interpretable outcomes, we provide an additional explanation based on the identified results, aiming to elucidate the process through which the attack behavior occurs.  Therefore, in the pattern analysis phase, the output not only identifies one or more vulnerabilities reflected by the event information but also analyzes and deconstructs the process by which these vulnerabilities arise.  This ensures that the event definitively aligns with the pattern of the identified vulnerability.  In Section 2, we provide a detailed explanation of four different types of vulnerabilities, specifically including reentrancy, integer overflow, flash loan attacks, and DoS, offering a comprehensive foundation for understanding their characteristics and behaviors.

\section{Experiment}

\subsection{Dataset}


We evaluated ETrace on four Etherscan-recorded attack events.  For each, we analyzed both event-level explanations and the overall judgment.  Accuracy was assessed by checking whether the outputs captured relevant attack behaviors and aligned with actual attack processes.

\subsection{Performance}


\begin{table*}[ht]
\centering
\renewcommand{\arraystretch}{1.3} 
\begin{tabular}{|>{\raggedright\arraybackslash}p{2cm}|
                >{\raggedright\arraybackslash}p{3cm}|
                >{\raggedright\arraybackslash}p{12cm}|} 
\hline
\textbf{Event} & \textbf{Actual Attack} & \textbf{Comprehensive Judgment} \\
\hline
XSURGE & \textbf{Reentrancy} & There are consecutive transfers between the same addresses (e.g., from address 0x5f2e to address 0xE1E1) with varying values. This could potentially be a \textbf{reentrancy risk} if the state is not updated correctly. \\
\hline
Beauty Chain & \textbf{Integer Overflow} & The large integer value is suspicious and could suggest potential \textbf{integer overflow}, especially in systems where such values exceed the maximum storage capacity for integers. Given that the value is extremely large, it should be examined for whether it exceeds the limits of the token's implementation. \\
\hline
MEVBOT & \textbf{Flash Loan Attack} & The FlashLoan event followed by the Swap events with abnormal price fluctuations could potentially indicate a \textbf{FlashLoanAttack event}. \\
\hline
GovernMental & \textbf{DoS} & The transaction with a value of 1.0 ETH and gas used of 750000 to an unknown function could potentially be a \textbf{DoS attack}. The abnormal gas used values in some transactions could suggest potential inefficiencies or issues in the contract code. \\
\hline
\end{tabular}
\caption{Compehensive Judgement}
\vspace{-2em}
\label{result}
\end{table*}

The result of the LLMs is shown in Table .\ref{result}.  Based on the results, LLMs are able to provide reasonable analysis and explanations of the vulnerabilities reflected by the events, including the causes of the attacks and the types of attacks.  The experimental results demonstrate that LLMs can not only effectively parse fine-grained event content, aiding in the understanding of smart contract behavior, but also integrate multiple event details to make comprehensive judgments.  Although the contextual relevance of LLMs may slightly decrease for longer event information due to token limitations, these results still validate the feasibility of our approach.

However, a limitation is that no threshold was set for the value, which caused LLMs to incorrectly identify integer overflows during the analysis of reentrancy events when the value was too high.  This is an aspect we plan to improve in the future.

\section{Related work}

\textbf{Analysis Tools.} Smart contract vulnerability detection is crucial, typically using three approaches: static analysis, dynamic analysis, and formal verification.
Static analysis tools exemplified by Oyente \cite{8} employ symbolic execution and pattern-matching algorithms to scrutinize contract bytecode/source code against predefined vulnerability signatures\cite{9}, effectively identifying risks such as reentrancy attacks and integer overflows. Dynamic analysis frameworks like Mythril \cite{10} and VERITE\cite{kong2025smart} adopt a complementary approach by executing contracts in sandboxed Ethereum Virtual Machine (EVM) environments, enabling real-time detection of runtime vulnerabilities through transaction simulation and state monitoring. Formal verification systems represented by Sereum \cite{11} leverage mathematical models to establish rigorous security guarantees, But their practical adoption faces challenges due to the significant manual effort for specification development and proof construction.

\textbf{LLMs Analysis.} As LLMs developed \cite{13}
, it was gradually used to replace some manual work. At the same time, the integration of LLMs into smart contract security \cite{15,16} represents a novel and promising direction.
However, LLMs exhibit hallucination tendencies, generating factually inconsistent or ungrounded outputs, which limits their applicability to unverified contracts.
Recent researches \cite{12,7} have revealed that events in smart contracts demonstrate their valuable significance. This empirical evidence motivates our exploration of LLMs' pattern reasoning capability with event-driven behavioral semantics to overcome hallucination limitations.

\section{Conclusion}
To address the lack of source code access, we propose ETrace, a framework that uses LLMs and event data from transaction logs to detect smart contract vulnerabilities. ETrace performs semantic analysis of events, enhances interpretability of contract behaviors, and identifies four types of abnormal activities. We validate its effectiveness on four real-world attacks.
In the future, we plan to expand ETrace by mining more event data to improve its versatility and accuracy.

\section{Acknowledgment}
This work was supported by National Key Research and Development Program of China (2022YFB2703500), and National Natural Science Foundation of China (62372367).



\bibliographystyle{unsrt}
\bibliography{ref}  
\end{document}